\documentclass[prl,twocolumn,showpacs,superscriptaddress,preprintnumbers]{revtex4}

\usepackage{graphicx}
\usepackage{dcolumn}
\usepackage{bm}
\usepackage{color} %

\begin{document}

\preprint{APS/123-QED}

\title{Versatile helimagnetic phases under magnetic fields in cubic perovskite SrFeO$_{3}$
}

\author{S. Ishiwata}
 \email{ishiwata@ap.t.u-tokyo.ac.jp}
 \affiliation{Department of Applied Physics and Quantum-Phase Electronics Center (QPEC), University of Tokyo, Hongo, Tokyo 113-8656, Japan}
 \affiliation{Cross-Correlated Materials Research Group (CMRG) and Correlated Electron Research Group (CERG), RIKEN-ASI, Wako 351-0198, Japan}

\author{M. Tokunaga}%
\affiliation{Institute for Solid State Physics, University of Tokyo, Kashiwa 277-8581, Japan}%

\author{Y. Kaneko}%
\affiliation{Multiferroics Project, ERATO, Japan Science and Technology (JST), c/o RIKEN, Wako 351-0198, Japan}%

\author{D. Okuyama}%
\affiliation{Cross-Correlated Materials Research Group (CMRG) and Correlated Electron Research Group (CERG), RIKEN-ASI, Wako 351-0198, Japan}%

\author{Y. Tokunaga}%
\affiliation{Cross-Correlated Materials Research Group (CMRG) and Correlated Electron Research Group (CERG), RIKEN-ASI, Wako 351-0198, Japan}
\affiliation{Multiferroics Project, ERATO, Japan Science and Technology (JST), c/o RIKEN, Wako 351-0198, Japan}%

\author{S. Wakimoto}%
\affiliation{Quantum Beam Science Directorate, Japan Atomic Energy Agency, Tokai, Ibaraki 319-1195, Japan}%

\author{K. Kakurai}%
\affiliation{Quantum Beam Science Directorate, Japan Atomic Energy Agency, Tokai, Ibaraki 319-1195, Japan}%

\author{T. Arima}%
\affiliation{Department of Advanced Materials Science, University of Tokyo, Kashiwa 277-8561, Japan}%

\author{Y. Taguchi}%
\affiliation{Cross-Correlated Materials Research Group (CMRG) and Correlated Electron Research Group (CERG), RIKEN-ASI, Wako 351-0198, Japan}%

\author{Y. Tokura}%
\affiliation{Department of Applied Physics and Quantum-Phase Electronics Center (QPEC), University of Tokyo, Hongo, Tokyo 113-8656, Japan}%
\affiliation{Cross-Correlated Materials Research Group (CMRG) and Correlated Electron Research Group (CERG), RIKEN-ASI, Wako 351-0198, Japan}%
\affiliation{Multiferroics Project, ERATO, Japan Science and Technology (JST), c/o RIKEN, Wako 351-0198, Japan}%

\date{\today}

\begin{abstract}
A helical spin texture is of great current interest for a host of novel spin-dependent transport phenomena. We report a rich variety of nontrivial helimagnetic phases in the simple cubic perovskite SrFeO$_{3}$ under magnetic fields up to 42 T. Magnetic and resistivity measurements revealed that the proper-screw spin phase proposed for SrFeO$_{3}$ can be subdivided into at least five kinds of ordered phases. Near the multicritical point, an unconventional anomalous Hall effect was found to show up and was interpreted as due to a possible long-period noncoplanar spin texture with scalar spin chirality.
\end{abstract}

\pacs{75.30.-m, 75.30.Kz, 72.80.Ga}
\maketitle

\section{Introduction}
 Topological spin textures have attracted considerable attention because of their potential for novel quantum transport phenomena and spintronic functions \cite{Day}. One such example is a skyrmion spin texture, the spin-swirling topological object, observed for B20-type compounds $MX$ ($M$ = Mn, Fe, Co; $X$ = Si, Ge) with a chiral cubic lattice \cite{Robler, Muhlbauer, Yu_Nature, Yu_NMat}. While Dzyaloshinskii-Moriya interaction in the noncentrosymmetric lattice yields the spin spiral in these compounds \cite{Dzyalo, Moriya}, effects of spin frustration can also be a source for a rich variety of helical spin textures with nearly degenerate free energies. Since the degeneracy remains intact in a high symmetry lattice, a cubic lattice with helical spin orders will be an ideal system hosting novel competing phases in the field-temperature phase plane. 

The compound SrFeO$_{3}$ investigated here is a rare oxide showing a helimagnetic order and metallic conduction while preserving the cubic lattice symmetry at least above the magnetic transition temperatures ($T >$ 130 K) \cite{MacChesney,Takeda_screw,Oda_screw}.  The x-ray absorption spectroscopy has revealed a negative charge transfer (CT) energy with strong $p-d$ hybridization, giving rise to an itinerant ligand hole $\underline{L}$ in the oxygen $p$ orbital \cite{Bocquet,Abbate}. Thus, the realistic electronic state of the Fe$^{4+}$ ions is rather close to $d^{5}\underline{L}$ than $d^{4}$, being free from the strong Jahn-Teller instability \cite{Bocquet}. The origin of the spiral spin order has been ascribed to the competition between the ferromagnetic nearest-neighbor interaction and the antiferromagnetic next-nearest-neighbor interaction \cite{Takeda_screw} or to the double-exchange mechanism working in the negative CT energy system \cite{Mostovoy}.

So far, the magnetic state of SrFeO$_{3}$ has been found not to be straightforward at all. While the proper-screw spin order with a propagation vector $\bm{q}$ along $\langle$111$\rangle$ below 134 K was confirmed by neutron diffraction experiments \cite{Takeda_screw, Oda_screw}, the occurrence of two more successive transitions near 110 K and 60 K has been reported  on the basis of resistivity and magnetization data \cite{Zhao, Lebon, Adler}. Near 60 K, where a first order transition occurs, large negative magnetoresistance (MR) under magnetic fields $H$ up to 9 T was reported for single crystals of SrFeO$_{3-x}$ ($x \leq$ 0.05). Hayashi $et$ $al.$ prepared a thin film of SrFeO$_{3}$ and found an anomalous enhancement of the Hall coefficient below 100 K \cite{Hayashi}. However, mainly due to the limitation of the field range, the perspective magnetic phase diagram as well as the interplay between the helimagnetic and the transport properties has remained to be explored. 

In this paper, we studied the magnetic phase diagram of SrFeO$_{3}$ by resistivity and magnetization measurements for the single crystalline sample under high $H$ up to 42 T, and we found versatile competing helimagnetic phases. Furthermore, unconventional Hall effect has been found near the cross point of the phase boundaries, i.e., the multicritical point. This feature is ascribed to a topological Hall effect as observed in noncoplanar spin texture with scalar spin chirality. When a conduction electron senses a long-period topological spin texture through Hund's-rule coupling, the Berry phase term is added to the wave function, yielding a fictitious magnetic field in the real space as the source of topological Hall effect \cite{Onoda, Binz, Neubauer, Nagaosa, Kanazawa}.

\begin{figure}[]
\includegraphics[keepaspectratio,width=6.5 cm]{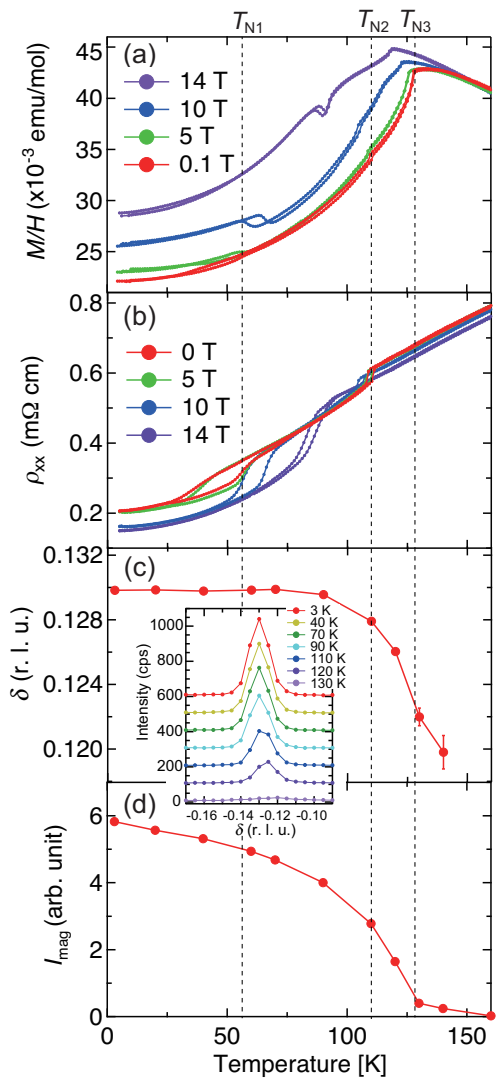}
\caption{\label{fig1} (Color online) Temperature dependence of (a) magnetization divided by external magnetic fields $H$ (0.1$-$14 T along $[$111$]$), (b) resistivity $\rho_{xx}$ with dc current perpendicular to $H$ (0$-$14 T along $[$111$]$), (c) $\delta$ component of the helimagnetic propagation vector, and (d) integrated intensity of the scattering peak at (1-$\it{\delta}$ 1-$\it{\delta}$ -$\it{\delta}$) for a single crystal of SrFeO$_{3}$. Representative peak profiles around (1-$\it{\delta}$ 1-$\it{\delta}$ -$\it{\delta}$) are shown in the inset. The unit "r. l. u." is the abbreviation of reciprocal lattice unit.
}
\end{figure}

\section{Experimental}
Single crystals of SrFeO$_{3}$ were obtained by a high-pressure treatment of single crystalline samples of SrFeO$_{2.5}$ with the brownmillerite structure,  which is the similar procedure adopted for the single-crystal growth of the isostructural compound SrCoO$_{3}$ \cite{Long}. The single crystal of SrFeO$_{2.5}$ was grown by a floating-zone method in an Ar gas flow. The obtained cylindrical crystal with a diameter of about 2.5 mm was cut to a suitable size for a gold capsule and then was heat treated with oxidizer KClO$_{4}$ for 1 h at 1073 K and 6.5 GPa. The crystallinity and the orientation of the samples were checked by x-ray Laue diffractions. Longitudinal resistivity $\rho_{xx}$, transverse  (Hall) resistivity $\rho_{yx}$, and magnetization $M$ were measured for magnetic fields $H$ up to 14 T by a physical property measurement system (Quantum Design) with a vibrating sample magnetometer. High $H$ measurements for $\rho_{xx}$ and $M$ up to 42 T were performed by using a nondestructive long-pulse magnet with a duration time of about 36 ms, installed at the Institute for Solid State Physics, University of Tokyo. 

\section{Results and discussion}
Figure 1 shows the thermal variation of the helical spin order in SrFeO$_{3}$ with successive phase transitions. Three anomalies at 56 K ($T_{\rm{N1}}$), 110 K ($T_{\rm{N2}}$), and 130 K ($T_{\rm{N3}}$) can be discerned in the heating runs of $M$ measured under $H$ of 0.1 T and $\rho_{xx}$ without $H$ (Figs. 1(a) and 1(b)). $T_{\rm{N3}}$ corresponds to the onset of the helical spin order as confirmed by the appearance of a magnetic satellite peak of (1-$\it{\delta}$ 1-$\it{\delta}$ -$\it{\delta}$) (see the inset of Fig. 1). The integrated intensity of the magnetic peak $I_{\rm{mag}}$ increases gradually and monotonously with decreasing temperature from $T_{\rm{N3}}$ down to the lowest temperature, as shown in Fig.1(d). However, no sign indicating further magnetic transitions nor spin reorientations was discernible at $T_{\rm{N1}}$ and $T_{\rm{N2}}$ in the temperature dependences of $I_{\rm{mag}}$ and $\delta$ or in the diffraction profiles scanned along $[\bar{1}\bar{1}\bar{1}]$ (Figs. 1(c) and 1(d)), perhaps due to the shortage of resolution in this neutron diffraction experiment. The $\delta$ value shown in Fig.1(c) increases from 0.12 at 130 K to 0.13 below 70 K, corresponding to the smooth change in the periodicity of the spin spiral from 18.5 $\rm{\AA}$ to 17 $\rm{\AA}$. As shown in Fig. 1(b), while the onset of the helimagnetic order at $T_{\rm{N3}}$ has little impact on the resistivity $\rho_{xx}$, the subsequent transitions cause clear drops in $\rho_{xx}$, in accord with the results from previous reports \cite{Hayashi, Zhao, Lebon, Adler}. When increasing $H$ up to 14 T, $T_{\rm{N1}}$ and $T_{\rm{N2}}$ get closer to each other and almost merge at around 90 K, suggesting the presence (at $H <$ 14 T) as well as the disappearance (at 14 T) of the ordered phase (phase II) in between $T_{\rm{N1}}$ and $T_{\rm{N2}}$. 

In order to see the $H$-dependent stability of the respective phases, we performed $M$ and $\rho_{xx}$ measurements under pulsed magnetic field ($H\|$ $I_{\rm{current}}$$\|$ $[$111$]$) at selected temperatures (Figs. 2 and 3). At 4 K, the $M-H$ curve shows a saturation of about 3.5 $\mu_{\rm_{B}}$ at 38 T, which is slightly smaller than the value expected for the high-spin state of $d^{5}\underline{L}$ ($S$ = 2) with the half-metallic (fully spin-polarized) structure. In addition to the $H$-induced ferromagnetic phase at low temperatures, five kinds of ordered phases I$\sim$V could be unraveled through the measurements of $M$ and $\rho_{xx}$ as functions of $H$ and $T$, as summarized in Fig. 4. The shaded area embedded in the phase I represents the region where $\rho_{xx}$ shows $H$-dependent hystereses after ZFC. Once after the drastic drop in $\rho_{xx}$ is induced by application of $H$, $\rho_{xx}$ never recovers to the initial value with the removal of $H$, however it does with heating above 60 K under zero field (see Fig. 3). We attributed this hysteretic behavior to the $H$-induced rotation of the helimagnetic domains which are originally degenerate with the $q$ vectors along one of the equivalent four $\langle$111$\rangle$ directions at zero field in the cubic symmetry, as reported for MnSi \cite{Grigoriev}. The application of high enough $H$ is expected to align all the $q$ vectors along one of them close to the $H$ direction, being consistent with the irreversible negative MR \cite{Zhao, Lebon, Adler}. 

\begin{figure} []
\includegraphics[keepaspectratio,width=5.8 cm]{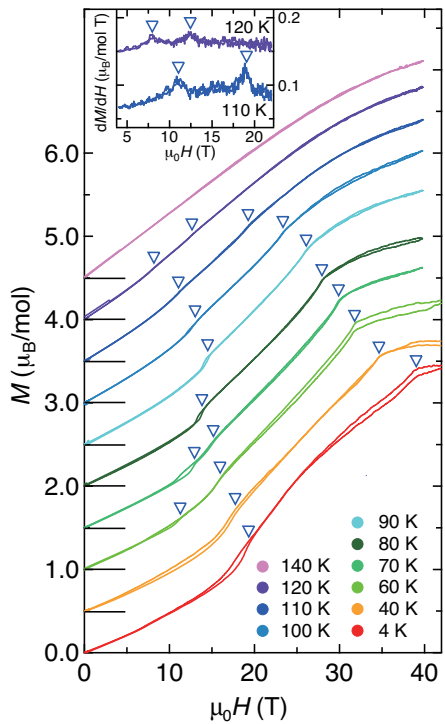}
\caption{\label{fig2} (Color online) Magnetic field dependence of magnetization along $[$111$]$ at selected temperatures. Magnetization curves except for the 4-K data are shifted upward for clarity. The open triangles represent the transition fields. The field derivatives of magnetization d$M$/d$H$ for the data taken at 110 K and 120 K are shown in the inset (the 120-K data are shifted upward for clarity.)
}
\end{figure}

\begin{figure} []
\includegraphics[keepaspectratio,width=5.8 cm]{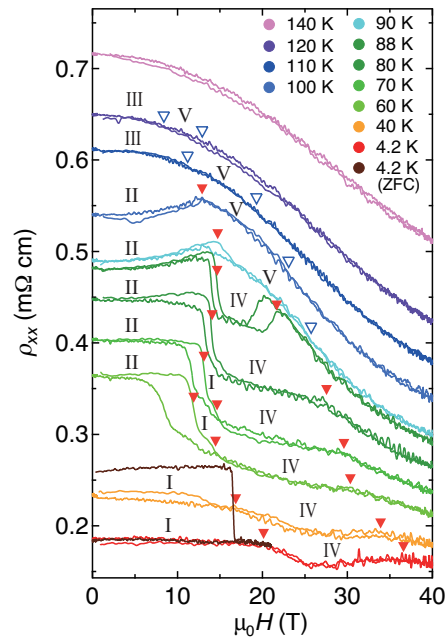}
\caption{\label{fig2} (Color online) Magnetic field dependence of resistivity at selected temperatures. The open and closed triangles represent the transition fields determined by the magnetization and resistivity data, respectively. For notations I$\sim$V, see Fig. 4. Pulsed high magnetic fields were applied along $[$111$]$ and the voltage was measured along the same direction. The resistivity at 4 K was measured after zero-field cooling (ZFC) under magnetic fields up to 20 T and then measured again at the same temperature under magnetic fields up to 40 T.
}
\end{figure}

\begin{figure}[]
\includegraphics[keepaspectratio,width= 8.5 cm]{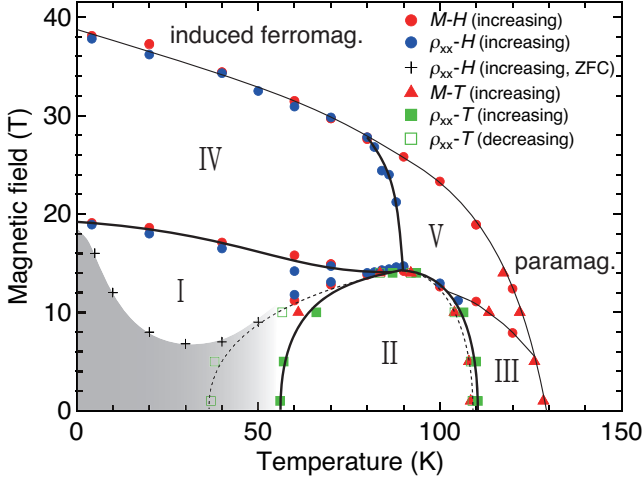}
\caption{\label{fig3} (Color online) Magnetic phase diagram of SrFeO$_{3}$. Magnetic field $H$ and the current direction are parallel to $[$111$]$. The solid lines separating phases I $\sim$V and the paramagnetic phase were determined by the $T$- and $H$-increasing runs of magnetization $M$ and resistivity $\rho_{xx}$ measurements. The thick and narrow lines indicate the first-order and second-order phase boundaries, respectively. The broken lines were determined by the $T$-decreasing run of $\rho_{xx}$. The shaded area was characterized as a hysteretic region associated with the $H$-induced irreversible flop of the helical modulation vector, as determined by the $H$ dependence of $\rho_{xx}$ measured after ZFC.} 

\end{figure}

Above 110 K, in phases III and V as well as in the paramagnetic phase, the negative MR defined by -$[\rho_{xx}$($H$)-$\rho_{xx}$(0)$]$/$\rho_{xx}$(0) is quadratic with $H$ (Fig. 3), while $M$ is almost linear with $H$ (Fig. 2). This behavior is characteristic of the spin scattering of the conduction electrons that is suppressed by the magnetic field. The negative MR at 40 T is as large as -30 \% in the temperature range involving these phases. In addition, the slope of $\rho_{xx}$ vs $H$ curves shows a gradual change without significant anomalies upon the transitions as III $\rightarrow$ V $\rightarrow$ paramagnetic phase. These results suggest that the spin fluctuation in phases III and V is significantly large, even comparable to the paramagnetic phase. In phase II, by contrast, a positive MR is observed, implying that the proper-screw like spin structure is deformed into a nontrivial spin texture by the application of $H$.

At 88 K, the reentrant behavior in MR accompanied with hystereses was observed in the $H$ range of 14$-$23 T, corresponding to the first-order transitions as II $\rightarrow$ IV $\rightarrow$ V. The application of $H$ transforms the proper-screw spin structure into the ferromagnetic one through an intermediate state such as conical or fan-like spin structure \cite{Nagamiya}. When $H$ is parallel to the screw axis, the proper-screw spin structure would continuously transform into conical spin structure. However, when $H$ is perpendicular to the screw axis, the fan-like spin structure is favored for the system with strong planar anisotropy, whereas the transverse conical spin structure with the cone axis parallel to $H$ becomes more stable in the absence of the strong anisotropy. In the case of SrFeO$_{3}$, the magnetic anisotropy is expected to be rather small, since the orbital moment in the $d^{5}\underline{L}$ state is quenched under the cubic crystal field. In fact, the saturation magnetic fields for all the directions of $\langle$111$\rangle$, $\langle$110$\rangle$, and $\langle$100$\rangle$ are almost identical (data not shown). Therefore, the fan-like spin structure can be ruled out and the conical spin structure would be the most likely candidate for the intermediate states such as phases IV and V. There are two possible states for the conical spin structures, the one having a single helimagnetic domain with $\bm{q}$ vector parallel to $H$ and the other having helimagnetic domains with multiple $\bm{q}$ vectors along the equivalent four $\langle$111$\rangle$ directions. From the viewpoint of entropy, the former and the latter are expected to be realized in phases IV and V, respectively. Thus, the increase in $\rho_{xx}$ on the transition from phase IV to V reflects the enhancement of the fluctuation of localized spins and/or the presence of the helimagnetic $q$ domains.

\begin{figure}[]
\includegraphics[keepaspectratio,width= 8.5 cm]{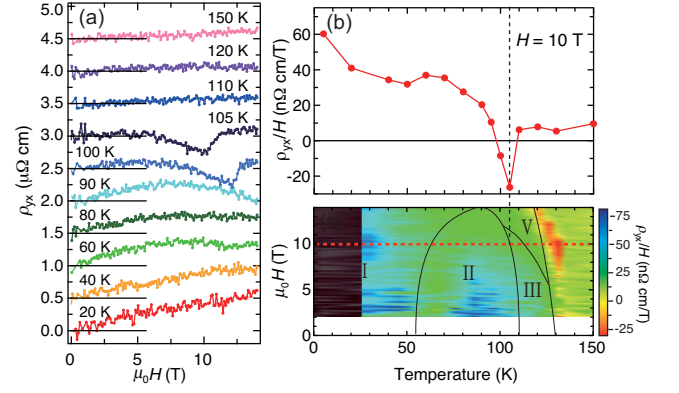}
\caption{\label{fig4} (Color online) (a) Magnetic field $H$ dependence of Hall resistivity $\rho_{yx}$ measured on increasing $H$ ($\|[$111$]$). The $\rho_{yx}$ curves except for the 20-K data are shifted upward for clarity. The electric current is parallel to $[$1$\bar{1}$0$]$. All the data were measured after field cooling at $H$ of 14 T. (b) Temperature dependence of $\rho_{yx}$ divided by $H$ of 10 T. The vertical broken line indicates the transition between phases II and III. (c) Contour map of $\rho_{yx}$ divided by $H$. The horizontal broken line corresponds to the trace shown in (b).
}
\end{figure}

Characterization of phases I and II has been performed through the Hall resistivity measurements. Figure 5(a) shows $\rho_{yx}$ as a function of $H$ at selected temperatures. Above 110 K involving phases III and V, $\rho_{yx}$ is linearly proportional to $H$, reflecting the anomalous Hall effect (AHE) proportional to $M$ that is almost proportional to $H$, in addition to the normal Hall effect (NHE) as the $p$-type conductor. Below 110 K down to 60 K, $\rho_{yx}$ shows nonlinear behavior with a steep decrease at around 10 T. Provided that the fictitious magnetic field $B_{\rm{eff}}^{z}$ in the real space contributes to $\rho_{yx}$ in addition to the conventional terms proportional to $H$ (NHE) and $M$ (AHE), $\rho_{yx}$ can be represented by the following equation,
\begin{equation}
 \rho_{yx} = \mu_{0}R_{0}H + \mu_{0}R_{s}M + R_{0}B_{\rm{eff}}^{z}
\end{equation}
Here, we assumed that conduction electrons coupled to the spin texture can sense the $B_{\rm{eff}}^{z}$, or equivalently the Berry phase. Since $M$ increases with increasing $H$, the negative contribution to $\rho_{yx}$ observed for phase II can be assigned to $R_{0}B_{\rm{eff}}^{z}$. Considering that there would be helimagnetic chiral domains, this result signifies that the topological Hall effect is induced by a helical spin texture with a multiple-$\bm{q}$ structure, such as the skyrmion lattice, where the scalar spin chirality is present but is not dependent on the swirling direction of the spins. 

In Fig. 5(c), a contour map of $\rho_{yx}/H$ is superimposed on the $H$-$T$ phase diagram of SrFeO$_{3}$, which highlights the negative contribution of  $R_{0}B_{eff}^{z}/H$. On the verge of phase II, near the multicritical point shared by phases II, III, and V, the negative $\rho_{yx}/H$ region suggesting the presence of the nontrivial, topological spin texture. The emergence of this state was clearly seen in the temperature dependence of $\rho_{yx}/H$ at 10 T (see Fig. 5(b)). As decreasing temperature, $\rho_{yx}/H$ drops dramatically upon the transition from phase III to II, followed by recovering to be positive and approaching a much larger value than that in the paramagnetic phase. 

\section{Summary}
To summarize, we have found at least five kinds of magnetically ordered phases in SrFeO$_{3}$ in the temperature-magnetic field plane, presumably all with helical spin modulations. By the Hall resistivity measurements, an unconventional behavior, being distinct from the spin-orbit-coupling induced AHE, has been found in phases I and II. Although the complex spin structures of phases I$\sim$III were interconnected to a proper-screw structure by the neutron diffraction measurements, phases I and II should have long-period and higher hierarchy order of helical spin modulations, albeit to be elucidated by further study. On the verge of phase II close to the multicritical point shared by phases III and V, we have found the clear imprint of the emergence of a topological spin texture in the Hall resistivity behavior. This marginal stability is analogous to the skyrmion crystal phase in the B20-type compound which is surrounded by the first-order phase boundaries shared by the conical and the paramagnetic phases \cite{Muhlbauer}. The high-$H$ phases IV and V are expected to have conical spin structure, and the enhanced thermal spin fluctuation has been suggested for the higher-$T$ phases III and V. The rich phase diagram in SrFeO$_{3}$ reflects the highly symmetric, cubic lattice hosting the magnetically degenerated phases; this may lead to the emergence of higher-hierarchy topological spin textures as manifested by the topological Hall effect. The cubic perovskite SrFeO$_{3}$  remains as the simplest but fertile playground for the study of the helimagnetism induced quantum transport.

\begin{acknowledgments}
The authors thank N. Nagaosa, Y. Onose, N. Kanazawa, and M. Takano for useful comments. This study was in part supported by Grant-in-Aid for Scientific Research on Priority Areas "Novel States of Matter Induced by Frustration" (Grant Nos. 19052004, 20046017, and 19052001) from the MEXT, and by Funding Program for World-Leading Innovative R\&D on Science and Technology (FIRST Program), Japan.
\end{acknowledgments}

\end{document}